\documentstyle[prd,aps,floats]{revtex}

\begin{document}

\input epsf

\renewcommand{\topfraction}{0.99}
\renewcommand{\bottomfraction}{0.99}

\twocolumn[\hsize\textwidth\columnwidth\hsize\csname
@twocolumnfalse\endcsname

\title{Inflationary perturbations from a potential with a step}
\author{Jennifer Adams$^1$, Bevan Cresswell$^1$ and Richard Easther$^2$}
\address{$^1$University of Canterbury, Private Bag 4800, Christchurch,
  New Zealand } 
\address{$^2$ Institute for Strings, Cosmology and Astroparticle Physics, 
Columbia University, New York, NY 10027.}
\date{\today} 
\maketitle
\begin{abstract}
We use a numerical code to compute the density perturbations generated
during an inflationary epoch which includes a spontaneous symmetry
breaking phase transition. A sharp step in the inflaton potential
generates $k$ dependent oscillations in the spectrum of primordial
density perturbations. The amplitude and extent in wavenumber of these
oscillations depends on both the magnitude and gradient of the step in
the inflaton potential. We show that observations of the cosmic
microwave background anisotropy place strong constraints on the step
parameters.

\end{abstract}

\pacs{}

\vskip2pc]

\section{Introduction}

A period of inflation in the primordial universe provides a causal
explanation for the existence of the large scale structure observed
during the present epoch (reviewed in \cite{lythlid}).  The simplest
and most natural form of the scalar density perturbation spectrum is
the scale invariant case, or ${\cal P}_{{\cal R}} \propto k^{n-1}$
with $n=1$, where ${\cal R}$ is the curvature perturbation.

In principle, inflationary models driven by a continuously evolving
scalar field have a scale dependent spectral index, which can be
calculated using the ``slow roll'' approximation \cite{riotto}. This
expresses $n$ as a function of the inflaton potential and its
derivatives at the instant a mode leaves the horizon during inflation.
The inflaton evolves slowly, so only a small piece of the potential is
``sampled'' by the large scale structure in the present universe,
ensuring that, if the underlying potential is smooth, $n$ is not
strongly scale dependent.

Potentials with a ``feature'' at the value of the inflaton when
perturbations corresponding to astrophysical scales in the present
universe left the horizon can produce a primordial perturbation
spectrum with significant.  However, the inflaton moves slowly and
fine tuning is needed to put the feature in exactly the right part of
the potential. Thus, while it is possible to construct inflationary
models with a scale dependent spectrum, these models are often
somewhat contrived.

However, arguing that using a feature in the inflaton potential to
generate a complicated spectrum requires fine tuning assumes that the
potential has just one feature, but is otherwise smooth.  Adding a
large number of features to the potential makes it far more likely
that a randomly chosen piece of the perturbation spectrum will exhibit
considerable scale dependence.  In particular, Adams {\it et al.\/}
\cite{adams} showed that a class of models derived from supergravity
theories naturally gives rise to inflaton potentials having a large
number of sudden (downward) steps. Each step corresponds to a symmetry
breaking phase transition in a field coupled to the inflaton, since
the mass changes suddenly when each transition occurs.  In the
scenario studied by Adams {\it et al.\/}, a spectral feature is
expected every 10-15 e-folds, so if this model had driven inflation it
is likely that one of these features would be visible in the spectrum
extracted from observations of large scale structure (LSS) and the
cosmic microwave background (CMB).

Motivated by the existence of models which naturally and generically
lead to scale dependent spectra, this paper carefully examines the
consequences of introducing a step in the inflaton potential.  We
focus on spectral features which may be observable in the large-scale
structure or cosmic microwave background anisotropy, and therefore had
their origin around 50 e-folds before the end of inflation.

We model the step by assuming the potential
\begin{equation} \label{potl}
V(\phi)=\frac{1}{2}m^2\phi^2 \left[1+ 
c\tanh\left(\frac{\phi-\phi_{\rm step}}{d}\right)\right], 
\end{equation}
for the inflaton field $\phi$. This potential has a step at
$\phi=\phi_{\rm step}$ with size and gradient governed by $c$ and $d$
respectively. For physically realistic models, inflation is not
interrupted but the effect on the density perturbations is still
significant.  If inflation is actually interrupted the effect on the
perturbation spectrum is severe enough to rule out models where this
happens during the interval of inflation corresponding to observable
scales.  In order to accurately evaluate the spectrum, we find that we
must evolve the evolution equations numerically, rather than relying
on the slow roll approximation.

Inflationary models with scale dependent spectral indices have been
examined in several previous investigations 
\cite{kofman,salopek,hodges1,hodges2,leach1,leach2}. In
particular, two recent papers, the first by Leach and Liddle
\cite{leach1} and the second by Leach {\em et al.\/} \cite{leach2},
rely, as we do, on numerical evaluations of the mode equation to
compute the density perturbation spectrum. Our analysis focuses on
small features in the potential. Conversely, \cite{leach1} and
\cite{leach2} examine potentials which produce very abrupt changes in
the inflationary dynamics, including the temporary cessation of
inflationary expansion, so the spectra discussed in \cite{leach1} and
\cite{leach2} are changed for all values of $k$ larger than some
critical value.  In contrast, the spectra we consider here are
essentially unchanged from their form at small $k$ once the
oscillations have died away. Moreover, most of the models discussed in
\cite{leach1} and \cite{leach2} would need to be carefully tuned in
order to produce observable effects in the spectrum\footnote{An
exception is the enhancement of the perturbations produced just before
the end of inflation, which might lead to the formation of primordial
black holes \cite{leach1}. These have observable consequences which
are distinct from observations of large scale structure.}, whereas the
mechanism described by Adams {\em et al.\/} can alter the observable
spectrum without fine tuning.

\section{Formalism}

In this section we reproduce the important equations governing the
evolution of scalar curvature perturbations and gravitational waves
during inflation.  We use the formalism developed by Stewart and Lyth
\cite{stewlyth} where the quantities of interest are the curvature
perturbation $\mathcal{R}$ and tensor perturbation $\psi$.

In the scalar case it is advantageous to define a gauge invariant
potential
\begin{equation}
u = -z \mathcal{R}
\end{equation}
where $z\equiv a\dot{\phi}/H$. We use the standard notation, where $a$
denotes the scale factor, $H$ the Hubble parameter, $\phi$ the inflaton
field, and a dot the derivative with respect to time $t$. 

The equation of motion for the Fourier components, $u_{\rm k}$, is
\begin{equation}  \label{uevln}
u_{\rm k}'' + \left( k^2-\frac{z''}{z}
\right)u_{\rm k}=0,
\end{equation}
where the prime denotes differentiation with respect to conformal time
and $k$ is the modulus of the wave number \cite{stewlyth,muk,muk2}.
The form of the solution depends on the relative sizes of $k^2$ and
$z''/z$. In the limit $k^2 \gg z''/z$, $u_k$ tends to the free field
solution
\begin{equation}  \label{freesc}
u_{{\rm k}} \rightarrow \frac{1}{\sqrt{2k}} \, e^{-ik\tau}\,,
\end{equation}
where the normalization is determined by the quantum origin of
the perturbations (see \cite{recon} for a more detailed discussion).
Conversely, in the limit $k^2 \ll z''/z$ the growing mode is
\begin{equation}
u_{{\rm k}} \propto z\
\label{eqn:regmode}
\end{equation}
which means that the curvature perturbation, 
\begin{equation}
|{\mathcal R}_{{\rm k}}|=|u_{{\rm k}}/z|,
\end{equation}
is constant in this regime.  The $z''/z$ term can be written as $2 a^2
H^2$ plus terms that are small during slow roll inflation, so that the
first regime applies to a mode well inside the horizon with $k \gg
aH$, and the second to super-horizon scales when $k \ll aH$.

The spectrum ${\cal P}_{{\cal R}}(k)$ is defined in the usual way
as 
\begin{equation}
\langle {\cal R}_{{\rm k}_1} {\cal R}^*_{{\rm k}_2} \rangle =
        \frac{2\pi^2}{k^3} {\cal P}_{{\cal R}} \delta^{3} \,
        (k_1- k_2) \,,
\end{equation}
and is given by
\begin{equation}
\label{pspec}
{\cal P}_{{\cal R}}^{1/2}(k) = \sqrt{\frac{k^3}{2\pi^2}} \, 
        \left| \frac{u_{{\rm k}}}{z} \right| \,.
\end{equation}

The mode equation for gravitational waves is 
\begin{equation}  \label{gwmode}
v_{\rm k}'' + \left( k^2-\frac{a''}{a}\right)v_{\rm k}=0,
\end{equation}
where $v_{\rm k}=a\psi_{\rm k}$. In slow roll inflation $a''/a \simeq
2 a^2 H^2$ and the behavior of $v_{\rm k}$ is again characterized by
whether the mode is inside or outside the horizon:
\begin{eqnarray}  \label{freegw}
v_{{\rm k}} &\rightarrow& \frac{1}{\sqrt{2k}} \, e^{-ik\tau}\, \quad {\rm
  as}\, \, aH/k \rightarrow 0, \\
v_{{\rm k}} &\propto& a \quad {\rm for} \, \, aH/k \gg 1.
\end{eqnarray}
The power spectrum of gravitational waves ${\cal P}_{g}(k)$
analogous to Eq. (\ref{pspec}) is
\begin{equation}
\label{gpspec}
{\cal P}_{g}^{1/2}(k) = \sqrt{\frac{k^3}{2\pi^2}} \, 
        \left| \frac{v_{\rm k}}{a} \right| \,.
\end{equation}

\section{Numerical Solution}

Normally, the perturbation spectra of inflationary models driven by a
continuously evolving, minimally coupled scalar field can be
calculated using the slow roll approximation. However, when the
potential has a sharp feature, its derivatives with respect to $\phi$
and the time derivatives of the field need not be small.
Consequently, we evolve the full mode equation numerically, without
any approximations other than those already implicit in the use of
perturbation theory.

In Eq.~(\ref{uevln}), the mode function is expressed in terms of
conformal time.  The intrinsic time-scale of the dynamics is not
constant in conformal time, so we shift the independent variable to
$\alpha = \log{a}$, facilitating the numerical integration.  With this
replacement, the system of equations we are to solve is
\begin{eqnarray}
&H_\alpha& = -4 \pi G H \phi_\alpha^2 \\
&\phi_{\alpha\alpha}& + \left(\frac{H_\alpha}{H} + 3 \right) \phi_\alpha 
+ \frac{1}{H^2} \frac{d V}{d\phi} =0 \\
&u_{\alpha\alpha}& + \left( \frac{H_\alpha}{H} + 1\right) u_\alpha 
+ \left\{ \frac{k^2}{e^{2\alpha} H^2} - \left[
2 -4 \frac{H_\alpha}{H}\frac{\phi_{\alpha\alpha}}{\phi_\alpha}
 \nonumber \right.\right.  \\
&& \left. \left. 
- 2 \left(\frac{H_\alpha}{H}\right)^2 -5\frac{H_\alpha}{H} -
\frac{1}{H^2} \frac{d^2 V}{d\phi^2}\right]\right\} =0
\end{eqnarray}
where the subscript $\alpha$ denotes differentiation. To compute the
spectrum, we repeat the integration for many values of $k$. 

In general, $u$ has two distinct solutions since it is a second order
linear differential equation, and we must choose the combination which
guarantees that the mode equation has the limiting form,
Eq.~(\ref{freesc}). We impose the initial conditions when the mode is
far inside the horizon assuming that the conformal time $\tau$ is
zero, which amounts to an irrelevant choice of phase. Consequently,
\begin{eqnarray}
\left. u \right|_{\tau=0} &=& \frac{1}{\sqrt{2k}}, \\
\left. \frac{d u}{d \alpha}\right|_{\tau=0} &=& -i 
\left. \sqrt{\frac{k}{2}} \frac{1}{e^\alpha H} \right|_{\tau=0}.
\end{eqnarray}
Rather than work with complex co-efficents in the numerical code, we
define two orthogonal solutions, $u^1_k$ and $u^2_k$, such that
\begin{eqnarray} \label{initcond1}
\left. u^1_k \right|_{\tau=0} &=& 1 \, , \\
\left. \frac{d u^1_k}{d\alpha} \right|_{\tau=0} &=& 0 \, , \\
\left. u^2_k \right|_{\tau=0} &=& 0 \, , \\
\left. \frac{d u^2_k}{d\alpha} \right|_{\tau=0} &=& 1 \, .\label{initcond4}
\end{eqnarray}
At any subsequent time $u_k$ is thus
\begin{equation}\label{comb}
u_k = \frac{1}{\sqrt{2k}} u^1_k - 
i\left. \sqrt{\frac{k}{2}}\frac{1}{e^\alpha H} \right|_{\tau=0} u^2_k.
\end{equation}

We start the evolution by evolving the two background equations until
any initial transient solution has died away but the mode is still
well inside the horizon. We then identify the two orthogonal solutions
that contribute to $u_k$ and extract the coefficients in (\ref{comb}).
This ensures that an initial transient contribution to the background
dynamics cannot contaminate the initial values of $u$ and
$u_\alpha$. Finally, to compute the spectrum, we need the asymptotic
value of $|u/z|$, and we find this by continuing the integration until
the mode is far outside the horizon and this value is effectively
constant.

The numerical integrations are carried out using the Bulirsch-Stoer
algorithm \cite{PressBK1}, and we check our calculations by
ensuring that the results are independent of the distance inside the
horizon where we apply the normalization, and the distance beyond the
horizon where we evaluate the asymptotic value of $|u/z|$.
\section{Inflationary potential with a step}

Figure 1 shows the power spectrum for the potential of
Eq. (\ref{potl}) with $c=0.002$, or a 0.4\% change in the amplitude of
the potential. The most striking aspect of the scalar spectrum is the
scale dependent oscillations. Even with this small change in the
amplitude of the inflaton potential the oscillations last for two
decades of $k$ and, at their peak, change the amplitude of the
spectrum by a factor of 3. We have set the position of the step so
that the scale where the oscillations begin, $k_{\rm low}$, is probed
by observations of the galaxy correlation function and the anisotropy
in the cosmic microwave background.

\begin{figure}[t]
\centering 
\leavevmode\epsfysize=6cm \epsfbox{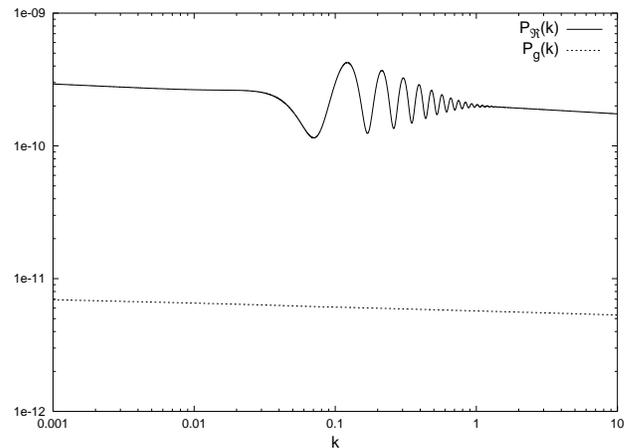}\\
\caption[fig1]{\label{fig:spec1} The scalar and tensor power
  spectrum for $c=0.002$ and $d=0.01$. The $z''/z$ term for these parameters is
  shown in Figure 2. }
\end{figure}

Before we look at the the origin of the oscillations in the scalar
spectrum it will be helpful to have a picture of how inflation
proceeds when there is a step in the potential. A general, qualitative
analysis of the spectrum produced by a ``feature'' in the potential is
given by Starobinsky \cite{staro}. For the specific model we are
considering here, we can understand the numerical results as
follows. Energy conservation requires that the change in the inflaton
kinetic energy term cannot exceed the change in the potential energy
so, if we are originally well inside the vacuum-dominated regime, a
small change in the amplitude of the inflaton potential cannot suspend
inflation. The evolution of $\ddot{a}$ in Figure 2 clearly shows that
the expansion is always accelerating.  However the $z''/z$ term, also
shown in Figure 2, determines the growth of the scalar perturbations
and is very different from $2a^2H^2$. It first grows in magnitude as
the inflaton field accelerates and then drops to a large negative
amplitude as the field slows. However, the tensor power spectrum is
unaffected since $a''/a$ remains constant throughout the step.

\begin{figure}[t]
\centering 
\leavevmode\epsfysize=6cm \epsfbox{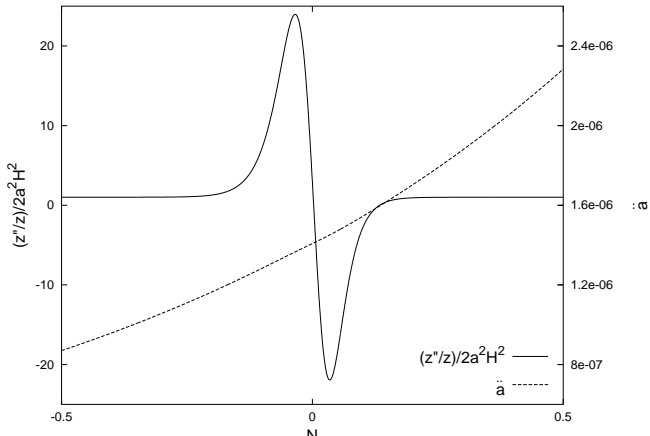}\\
\caption[fig2]{\label{fig:z''} Evolution of $z''/z$ and $\ddot{a}$ for
  $c=0.02$ and $d=0.01$ with the number of $e$-folds of inflation, $N$.
  We have set $N=0$ at the step in the potential.}
\end{figure}

\begin{figure}[t]
\centering 
\leavevmode\epsfysize=6cm \epsfbox{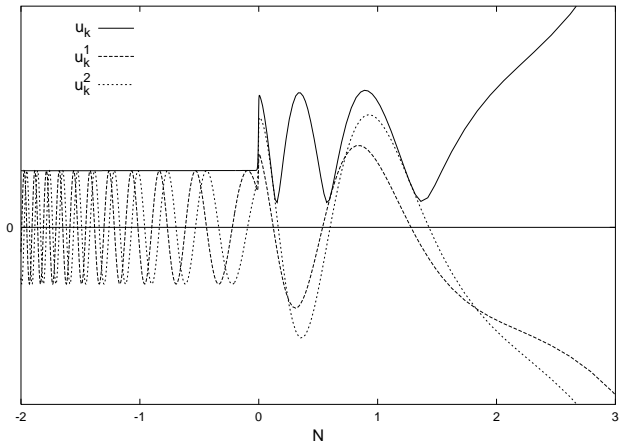}\\
\caption[fig3]{\label{fig:modes} Evolution of the independent modes
  $u^1_k$ and $u^2_k$ (with initial conditions for $u^1_k$ and $u^2_k$ given in Eq.(\ref{initcond1}-\ref{initcond4})) and
 the linear combination of their amplitude, Eq. (\ref{comb}) for $k=0.3$.}
\end{figure}

\begin{figure}[t]
\centering 
\leavevmode\epsfysize=6cm \epsfbox{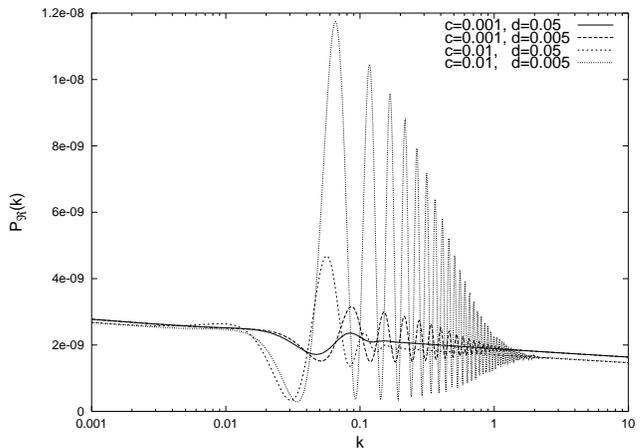}\\
\caption[fig4]{\label{fig:changingpar} The effect of changing $c$ and
  $d$ on the scalar perturbation spectrum.} 
\end{figure}

To understand the scalar power spectrum we begin by considering the
evolution of a particular scalar mode.  The evolution is governed by
the competition between the $k^2$ and $z''/z$ terms.  A step in the
potential of the magnitude we are interested in only has a lasting
effect on $k$ modes within the horizon, and not on modes which are
already well outside the horizon. That is the lowest wavenumber
affected is approximately given by $k_{\rm low} \sim aH|_{\rm step}$.
Moreover, from the form of the $k^2 - z''/z$ term in the mode
equation, we can see that the range of $k$ affected by the step will
scale roughly with the square root of the maximum value of $z''/z$ in
the region of the step.

In Figure 3 we show the evolution of $u^1_k$, $u^2_k$ and $u_k$ 
for an intermediate wavenumber in the range of $k$ affected. The rise
in $z''/z$ introduces a brief interlude of growing mode behavior into
the oscillatory regime. The subsequent interval where $z''/z$ is
negative causes the amplitude of $u_k$ to briefly resume its
oscillatory behavior. Finally, when the inflaton field resumes slow
rolling, the oscillations leave the horizon with altered phase and
increased amplitude. Both of the two initially independent solutions
are affected similarly, as they now have the same phase, and the
amplitude of their linear combination oscillates. In other words, the
presence of the step introduces a boundary condition which selects a
solution with an oscillating envelope in contrast to the unconstrained
plane wave solution with constant envelope seen at small $k$.

As in the case of a featureless inflaton potential $u_k$ obtains a
growing mode solution once it is outside the horizon. However the
asymptotic limit reached by the curvature perturbation $|{\mathcal
R}_{\rm k}|$ depends on the oscillation phase of the mode at horizon
exit, so that $|{\mathcal R}_{\rm k}|$ oscillates, with maxima
corresponding to the modes which exit at an extremum. The proper time
interval between the step and when the mode with wavenumber $k$ exits
is approximately $\Delta \tau \sim 1/aH|_{\rm step} - 1/aH|_{\rm exit}
= 1/k_{\rm low} - 1/k$ and in this time the amplitude of the mode will
have undergone $1/\pi(k/k_{\rm low}-1)$ oscillations. Thus the period
of the variation in $|{\mathcal R}_{\rm k}|$ is approximately $\pi
k_{\rm low}$.

For higher wavenumbers the effect of the $z''/z$ term is
smaller, the amplitude of their oscillation is not increased and the
two modes are not set exactly in phase with each other. However their
phases are still altered so that the linear combination $u_{\rm k}$
oscillates, but with a diminished amplitude compared to the lower $k$
modes.

The magnitude of $z''/z$ depends on both the parameters $c$ and $d$ in
the potential, and in a well motivated model these will be determined
by from particle physics.  Alternatively, given accurate observations
of the CMB and LSS, it may is possible cosmological constraints on the
values of these parameters, and in the next section we examine the
observable consequences of a scale dependent primordial
spectrum.

\section{Observable Spectra}

Adams {\it et al.} \cite{adams} attempted to recover the primordial
perturbation spectrum from the APM survey power spectrum using the
relationship between the spectrum of mass fluctuations today and the
primordial spectrum
\begin{equation}
{\cal P}_{\delta}\equiv {\cal P}_{{\cal
R}}T^2(k)\left(\frac{k}{H_0}\right)^{3+n}\, ,
\end{equation}
where $T(k)$ is the matter transfer function that tracks the
scale dependent rate of growth of linear perturbations and depends on
the dark matter content of the Universe. Assuming a CDM dominated
Universe the primordial $n(k)$ could be extracted from the three
dimensional $P_{\rm APM}$(k) inferred from the angular correlation
function of galaxies in the APM survey \cite{baughef}. A departure
from scale invariance was found in the range $k \sim (0.05-0.6) h
Mpc^{-1}$. This feature has been noted by \cite{baughef} and in the
power spectrum of IRAS galaxies \cite{peacock}. Adams {\it et al} used
the $n(k)$ they had extracted to predict the photon power spectrum and
found that the height of the secondary acoustic peaks was suppressed
by a factor of $\sim 2$.

Recently, a number of groups have revisited this analysis motivated by
the recent Maxima and BOOMERanG observations which show an anomalously
low second peak. Barriga {\it et al} performed a $\chi^2$ analysis of
the COBE and BOOMERanG CMB data considering a simple step in the
primordial perturbation spectrum (no oscillations).
Griffiths {\it et al} added a Gaussian bump to the primordial spectrum
and performed a similar exercise. Both groups found support for a
spectral feature.

\begin{figure}[t]
\centering 
\leavevmode\epsfysize=6cm \epsfbox{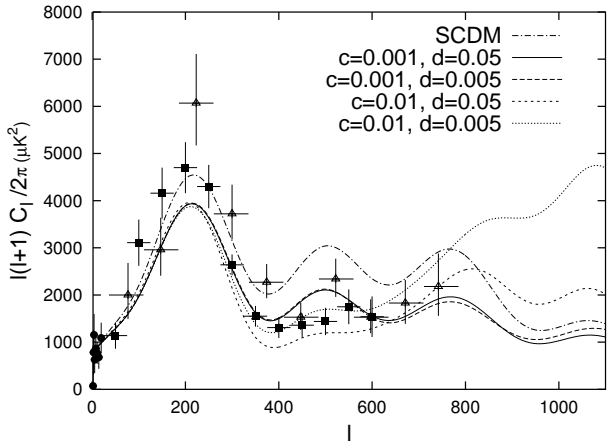}\\
\caption[fig5]{\label{fig:cmb} The CMB angular power spectrum for the
  primordial spectra shown in Figure 4. The normalization in each case
  is to COBE. The data are from COBE (circles), BOOMERanG (squares) and
MAXIMA (triangles).}
\end{figure}

We leave a $\chi^2$ analysis for a forthcoming paper and
include here for orientation the cosmic microwave background power
spectrum and matter power spectrum predicted for the range of primordial 
spectra shown in Figure 4. We use the Boltzmann code CMBFAST
\cite{seljak} to calculate the CMB angular power spectrum. We use the
less fashionable sCDM as our background cosmology ($\Omega_{\rm CDM} =
0.95,\,\Omega_{\rm B}=0.05, \, h=0.5 \,$) as our motivation is to
show the effect of the primordial density perturbation oscillations
rather than find the best fit. The CMB angular power spectrum is shown
in Figure 5 along with observations from COBE (the uncorrelated COBE DMR 
points from \cite{tegmark}), BOOMERanG \cite{boomerang} and MAXIMA
\cite{maxima}. It is clear that a good fit to the data can be found, but
that the amplitude of the step and its gradient are constrained to be
small by the observations.

\begin{figure}[t]
\centering 
\leavevmode\epsfysize=6cm \epsfbox{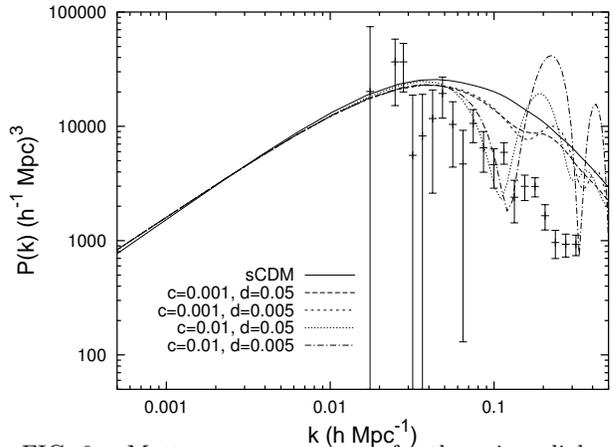}\\
\caption[fig6]{\label{fig:matter} Matter  power spectrum for the
  primordial spectra shown in Figure 4. The data points are from the
  PSCz catalogue.}
\end{figure}

The matter power spectrum is shown in Figure 6 along with the linear
power spectrum generated from the PSCz catalogue \cite{hamilton}. The
theoretical spectra suffer from the sCDM problem of too much power on
small scales however these spectra are given for indicative purposes
and are by no mean best-fit spectra.

\section {Conclusions}

We use a numerical routine to accurately calculate the primordial
density spectrum predicted by a {\it physically motivated\/} inflaton
potential with steps in it due to symmetry breaking during inflation.
The step in the potential induces oscillations in the density
perturbation spectrum whose magnitude and extent is dependent on the
amplitude and gradient of the step.

We have restricted our attention to a generic step, demonstrating that
even a small feature in the potential can cause significant changes to
the spectrum of large scale perturbations. Moreover, we have presented
a detailed account of how a feature in the potential modifies the
observable spectrum.  In the light of our calculations, we believe
that tight cosmological constraints can be placed on the size of any
feature in the potential, and thus on the particle physics model which
produced it, and intend to return to this problem in future work.
Conversely, if the density perturbation spectrum extracted from future
measurements of large scale structure and the cosmic microwave
background turns out to be incompatible with a smooth initial
spectrum, the mechanism proposed by Adams {\it et al.} provides a
natural mechanism for injecting significant scale dependence within
the context of inflation.
\cite{adams}

\section*{Acknowledgments} RE is supported by the Columbia University 
Academic Quality Fund, and thanks the University of Canterbury for its
hospitality while this work was begun.

\end{document}